\documentclass[draft]{agujournal2019}
\usepackage{url} 
\usepackage[inline]{trackchanges} 
\usepackage{soul}
\usepackage{graphicx}
\usepackage[export]{adjustbox}
\usepackage{wrapfig}
\usepackage{array} 
\usepackage{booktabs}
\usepackage{tablefootnote}
\usepackage{amsmath}

\draftfalse

\journalname{Geophysical Research Letters}

\begin{document}
%
%


\title{Electron Temperature Gradients Regulate the Duration of Two-Stage Plasmasphere Refilling}

%
%




\authors{Jaden Fitzpatrick\affil{1}, Kausik Chatterjee\affil{1,2,3}, Naomi Maruyama\affil{1}, Xiangning Chu\affil{1}, Jacob Bortnik\affil{4}, Jerry Goldstein\affil{5}, Lauren Blum\affil{1}, Tyler Bishop\affil{1}}

\affiliation{1}{Laboratory for Atmospheric and Space Physics, University of Colorado Boulder, Boulder, CO, USA}
\affiliation{2}{IB3 Global Solutions, Washington, DC, USA}
\affiliation{3}{University of New Mexico, Albuquerque, NM, USA}
\affiliation{4}{University of California Los Angeles, Los Angeles, CA, USA}
\affiliation{5}{Southwest Research Institute, San Antonio, TX, USA}




\correspondingauthor{Jaden Fitzpatrick}{jaden.fitzpatrick@lasp.colorado.edu}




\begin{keypoints}
\item Model simulations show how the field-aligned electron temperature, and thus gradient, regulate the early and late refilling stage durations
\item Second-degree multivariate regressions from the simulation results demonstrate the temperature's regulation of early and late stage lengths
\item Temperature structure may explain why some events exhibit only one observable refilling stage
\end{keypoints}

%
%

%
%

\begin{abstract}
    After geomagnetic storms erode the plasmasphere, cold ionospheric plasma flows along magnetic field lines to refill the depleted flux tubes. Previous studies have suggested that refilling undergoes two stages characterized by their distinct refilling rates; however, not all observations or models exhibit this feature, and a recent analysis indicates that only a subset of refilling events display two clear stages. In this study, we show that the magnitude of the field-aligned electron temperature gradient and initial/boundary temperature regulate the durations of both the early and late stages. Performing multivariate regressions from the results of simulating early and late-time refilling demonstrated that the length of each stage uniquely depends on the temperature gradient and initial/boundary temperature. These results suggest that variations in the temperature profile may explain why some refilling events appear single-staged in observations.
\end{abstract}

\section*{Plain Language Summary}

One result of solar storms is the erosion of the plasmasphere, which is a toroidal (doughnut-shaped), dense, and cold region of plasma that surrounds the Earth in space. After such erosion events, the plasma is replenished from the region below, the ionosphere. In this refilling process, ionospheric plasma travels upward along magnetic field lines to fill the emptied plasmasphere. Previous studies have suggested that refilling happens in two stages, each characterized by a distinct refilling rate. However, not all observations or models exhibit this feature, and a recent analysis indicates that only a subset of refilling events display two clear stages. In this study, we show that the electron temperature and the rate at which the electron temperature increases along the field line regulate the durations of both the early and late stages. These results suggest that variations in temperature with respect to altitude may explain why some refilling events appear single-staged in observations.


%
%

%


%
%
%
%

\section{Introduction}\label{Section1}

    Geomagnetic storms impact the Earth's magnetosphere, including the erosion of the plasmasphere - a torus-shaped reservoir of cold, dense plasma forming its innermost region. The plasmasphere is dynamically coupled to the ionosphere, allowing cold plasma from the ionosphere to flow upward along a magnetic field line and refill depleted flux tubes following a geomagnetic storm. Despite decades of research, the detailed processes governing plasmasphere refilling remain incompletely understood. Observations are complicated by the fact that plasmaspheric flux tubes do not strictly co-rotate with Earth, making refilling challenging to observe \cite{burch_cause_2004}. Furthermore, the hypothesized complex combination of processes involved make refilling difficult to model \cite{banks_dynamical_1971, khazanov_simulation_1984, singh_temporal_1986, rasmussen_multistream_1988, guiter_twostream_1995, wilson_semikinetic_1992, lin_semikinetic_1992, liemohn_nonlinear_1999}. A more thorough understanding of dynamic recovery of cold plasma populations would advance our knowledge of cross-energy interactions via wave-particle interactions \cite{usanova_role_2025} and enhance the accuracy of space weather predictions \cite{singh_state_2011}.\par
    While continuous observation of a single plasmaspheric flux tube throughout refilling would be ideal, early observational studies instead relied on measurements from geosynchronous orbit. Data from ESA's GEOS-2 satellite \cite{sojka_refilling_1985, song_refilling_1988} and from the Los Alamos National Laboratory (LANL) Magnetospheric Plasma Analyzers or MPAs \cite{lawrence_measurements_1999, su_comprehensive_2001} provided the first quantitative estimates of refilling rates. However, among these, only \citeA{lawrence_measurements_1999} and \citeA{su_comprehensive_2001} identified a two-stage refilling process. The first stage exhibited a low refilling rate compared to that in the second stage, and each stage was also attributed to a different physical mechanism: wave-particle interactions and Coulomb collisions, respectively. \par
    More recently, a comprehensive survey using the Van Allen Probes found that approximately $40\%$ of refilling events exhibited two stages \cite{bishop_superposed_2025}.  Due to a local time dependence in their distribution of two stage refilling observations, they found the occurrence of two stage refilling depended on the local plasma environment rather than global geomagnetic conditions like the DST index. The factors controlling whether two stages manifest or not remain uncertain. \par
    Given the observational limitations, substantial modeling efforts have sought to explain plasmasphere refilling. Early one-dimensional, single-stream hydrodynamic models \cite{banks_dynamical_1971, khazanov_simulation_1984, singh_temporal_1986} produced equatorial shocks later shown to be numerical artifacts \cite{rasmussen_multistream_1988}. Subsequent multistream formulations \cite{guiter_twostream_1995} improved the treatment of counterstreaming flows. The first semikinetic approaches \cite{wilson_semikinetic_1992, lin_semikinetic_1992} successfully reproduced counterstreaming without shocks, and were the first to suggest a distinct two-staged plasmasphere refilling process. The modeled stage relative durations and rates were later supported by observations \cite{lawrence_measurements_1999}. Each stage was thought to arise from plasma trapping by acoustic waves and Coulomb collisions \cite{wilson_semikinetic_1992, lin_semikinetic_1992}. Variations and refinements of these models \cite{liemohn_nonlinear_1999, krall_sami3_2013,chatterjee_semikinetic_2020} continue to advance our understanding of the refilling process. \par
    
    The model examined in this study \cite{chatterjee_multiion_2019, chatterjee_development_2020} is a one-dimensional, hydrodynamic, multistream, multi-ion, and multi-neutral treatment, explained in detail in Section \ref{Section 2}. This model was recently improved by \citeA{fitzpatrick_enhanced_2026} by incorporating the electron energy equation \cite{khazanov_analytic_1992}, using the Crank-Nicolson method \cite{smith_numerical_1985}. \citeA{schunk_electron_1978} lays out the observations and theoretical framework for electron temperature and heating rates in the F-region of the ionosphere, suggesting the expected conditions at the lowest altitudes of the plasmasphere. \citeA{khazanov_analytic_1992} adjusted the treatment of electron heating to be suited to the rest of the plasmasphere. But heating mechanisms throughout the plasmasphere, like photoelectrons, ring current particles, ion cyclotron radiation, and joule heating \cite{khazanov_analytic_1992}, all have spatiotemporal variations throughout refilling. Consequently, the effects of each of these mechanisms are combined into a single universal heating rate as in \citeA{khazanov_analytic_1992} and \citeA{rees_observations_1975}. The electron energy equation can be used to calculate the resulting electron temperature, but is also dependent on parameters like density that drastically change during refilling. But with a refilling model like \citeA{fitzpatrick_enhanced_2026}'s, the effects of electron heating rate and temperature on the refilling process can be quantified. \par
    
    The purpose of this study is to determine how the electron-temperature gradient, established by different combinations of initial temperature ($T_0$) and heating rate ($Q_e$), affects the durations of the early and late stages of plasmasphere refilling. As described in Section \ref{Section 2}, a set of controlled simulations is performed to vary these parameters. The results, presented in Section \ref{Section 3}, demonstrate that the magnitude of the field-aligned temperature gradient and the initial/boundary temperature regulate the length of each refilling stage. Cases with sufficiently small gradients produce very short early-time refilling, potentially explaining why some observed events appear to exhibit only a single refilling stage.

\section{Methodology} \label{Section 2}

    The hydrodynamic plasmasphere model used in this study \cite{chatterjee_multiion_2019,chatterjee_development_2020}  describes the transport of ions (H$^+$, He$^+$, and O$^+$ ) and neutrals (O and H) along a geomagnetic flux tube by solving the coupled continuity and momentum equations for each ion species. While the analogous equations for electrons are ignored, the electron population is assumed to have a density to maintain quasi-neutrality and enough collisions to form a thermal, Maxwellian energy distribution. The model employs a two-stream formulation, with plasma flows originating from the northern and southern hemispheres. An all-encompassing heating rate $Q_e$ is designated to account for all relevant plasmasphere heating mechanisms, such as photoelectrons, ring current ions, ion cyclotron waves, and joule heating \cite{khazanov_analytic_1992}. Earlier versions of the model \cite{chatterjee_multiion_2019, chatterjee_development_2020} assumed a spatially and temporally constant temperature; here, the temperature is allowed to vary both in space and time. \par

The goal for eliminating the previously assumed constant temperature \cite{chatterjee_multiion_2019, chatterjee_development_2020} was to produce more realistic refilling results whilst maintaining select simplifying assumptions to specifically isolate the effects of temperature variability. The analysis conducted by \citeA{fitzpatrick_enhanced_2026} demonstrated that the relative refilling rates, stage lengths, and ion compositions remained comparable to previous models and observations after implementing space- and time-dependent temperature. The temperature is also such that it is self-consistent, guaranteeing an appropriate relationship between temperature and ion concentrations for the duration of a simulation. As shown by Eq. 3 of \citeA{fitzpatrick_enhanced_2026}, the ambipolar electric field depends on the spatial distribution of temperature and ion concentrations. This implies that the new version of the model more accurately accounts for the ambipolar electric field and the consequential upward force along the flux tube, thus enhancing the modeled refilling rates.  \par

To avoid falsely attributing the improved refilling characteristics to sources other than temperature variability, the following assumptions were kept from previous model iterations. The temperature under consideration is that of both the electrons and ions, thus assuming that the temperature of each are equivalent. The value of $T_0$ set also becomes the constant boundary temperature for the entire simulation. Similarly, $Q_e$ is varied between simulations but is constant for all latitudes at all times. To still reflect as realistic refilling conditions as possible, the selection of values for $T_0$ and $Q_e$ are done carefully and described later in this section. Incrementally resolving these assumptions is the subject of future work. \par

Having provided the reasoning behind the chosen assumptions, a brief explanation of how temperature is calculated follows. A more comprehensive description can be found in \citeA{fitzpatrick_enhanced_2026}.  Again, the initial temperature along the flux tube is still assumed to be constant ($T_0$) with altitude but is altered across simulations. Once the simulation begins, only the flux tube boundaries remain at $T_0$ for the duration of the simulation. The temperature at all other intermediate latitudes are determined by solving the electron energy equation from \citeA{khazanov_analytic_1992}:

\begin{equation}
\frac{3}{2} k\,n_{e}\,A(s)\,\frac{\partial T_{e}}{\partial t}
  = \frac{\partial}{\partial s}
    \biggl[ A(s)\,K_{e}\,T_{e}^{5/2}\frac{\partial T_{e}}{\partial s} \biggr]
    + A(s)\,Q_{e} ,
\label{eq:heat}
\end{equation}
where $n_{e}$ is the electron density obtained from quasi-neutrality, $A(s)$ is the magnetic-flux-tube cross-sectional area, and $K_e$  is the electron thermal conductivity. A description of $K_e$ is provided in \citeA{fitzpatrick_enhanced_2026}, with the original expression of $K_e$ is from \citeA{schunk_transport_1970} and derived by \citeA{banks_charged_1966}. As previously mentioned, the value of $Q_e$ can vary between simulations but is assumed to be constant for all latitudes and time. The Crank-Nicolson method \cite{smith_numerical_1985} is then used to solve Eq. 1 at each time step for each point along the flux tube.


    Now that the model permits temperature to fluctuate across time and space, this study will evaluate the effects of changing $Q_e$ and $T_0$. The selected combinations of $Q_e$ and $T_0$ were based on values in previous literature. \citeA{rees_observations_1975} and \citeA{schunk_electron_1978} provide relevant heating rates and electron temperatures for the ionosphere at high altitudes based on observed and theoretical values. \citeA{khazanov_analytic_1992} summarizes these and other works to give an estimated heating rate for each of the primary ionosphere/plasmasphere heating mechanisms and electron temperatures corresponding to different total heating rates. Ultimately, the simulations presented in detail in the next section had values of $Q_e$ ranging from approximately $10^9$ to $10^{11}$ eV cm$^{-2}$ s$^{-1}$ and $T_0$ in intervals of $1000$K from $3000$K to $6000$K. To avoid choosing combinations of $Q_e$ and $T_0$ from the above values that induced unrealistic temperature profiles \cite{rees_observations_1975, comfort_thermal_1996}, the only results analyzed are from the 19 simulations where the absolute temperature difference with respect to altitude was at least 2000K . \par

    The initial conditions used for all remaining parameters in all simulations will reflect those chosen in \citeA{fitzpatrick_enhanced_2026}. The depleted flux tube under consideration by this model study is at L=4 and extends to $\pm 56^{\circ}$ in latitude and 1,500 km in altitude. The initial concentrations of each ion across latitude are displayed in Fig. \ref{SampleRefilling} of \citeA{fitzpatrick_enhanced_2026}. The relative amount of each approximately reflects equinox conditions during solar maximum. \par

    By altering only $Q_e$ and $T_0$ in the simulations evaluated by this paper, the variations in the resulting refilling events should be explainable by heating and overall temperature effects. Realistically, a variety of heating mechanisms \cite{schunk_electron_1978, rees_observations_1975, khazanov_analytic_1992} influence the heating rate, initial electron temperature, and boundary electron temperature. These are instead designated by $Q_e$ and $T_0$ and ignore spatiotemporal variations of any heating mechanisms. However, these simplifications allow for the resulting refilling characteristics, such as the length of each refilling stage, to be more directly compared with the corresponding $Q_e$ and $T_0$, as is done in the following section.

\section{Results \& Discussion} \label{Section 3}

\begin{figure}
\includegraphics[center]{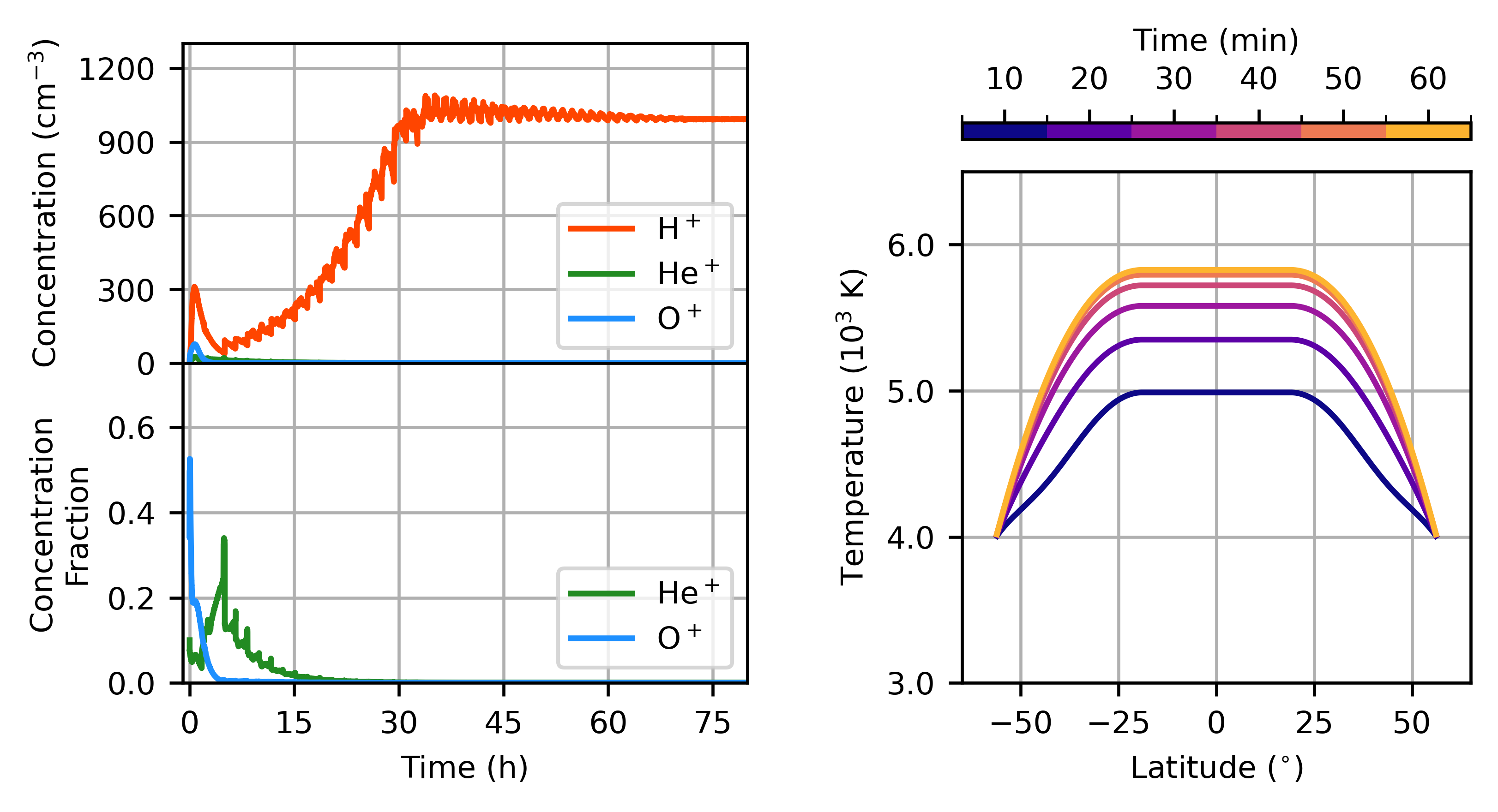}
\caption{(Top, Left) Equatorial concentration of H$^+$, He$^+$, and O$^+$ over time. (Bottom, Left) The fraction of equatorial concentration composed of He$^+$ and O$^+$. (Right) Evolution of the electron temperature profile across latitude and time, with each colored line corresponding to a time designated by the color bar.}
\label{SampleRefilling}
\end{figure}

    To become acquainted with the refilling profiles produced by the model used in this study, a detailed description of a single refilling result follows. Fig. \ref{SampleRefilling} was produced using the standard initial conditions assumed for all simulations just described in the previous section, as well as setting $T_0$ = 4000 K and $Q_e= 7 \times 10^9 $ eV cm$^{-2}$ s$^{-1}$. The left panel of Fig. \ref{SampleRefilling} shows the equatorial concentration of each ion during refilling. The initial peak nearly one hour into the simulation is due to the initial crossing of ions at the equator into the opposing hemisphere. The early stage of refilling is assumed to last from the start of the simulation until the H$^+$ concentration increases for the second time at approximately 5 h, in which case refilling has transitioned to the late stage. Also at the time of the stage transition is a maximum in the fractional concentration of He$^+$, which is thought to be facilitated by the ambipolar electric field and is explored further in \citeA{fitzpatrick_enhanced_2026}. Late-time refilling ends once an equilibrium concentration is reached, which for these conditions is approximately 45 hours after refilling began. The relative enhancement of total ion concentration after each refilling stage is consistent with previous modeling \cite{wilson_semikinetic_1992} and observations \cite{lawrence_measurements_1999, su_comprehensive_2001, bishop_superposed_2025}  where upon the conclusion of early-time refilling, only a marginally higher concentration is reached and during late-time refilling, a considerably larger refilling rate is observed. \par

    In the right panel of Fig. \ref{SampleRefilling} is the time progression of temperature profiles across latitude. While the temperature is initially constant across latitude, only one hour is necessary to reach a temperature profile nearly identical to the profile present for the remaining hours of refilling. The equilibrium temperature being reached so rapidly is a consequence of ignoring time-dependent heating like diurnal variations and enhancements from geomagnetic activity. Moreover, the model forgoes any loss terms as prescribed by \citeA{khazanov_analytic_1992}. These parameters will be considered in future model development. \par

\begin{sidewaysfigure}
\includegraphics[center]{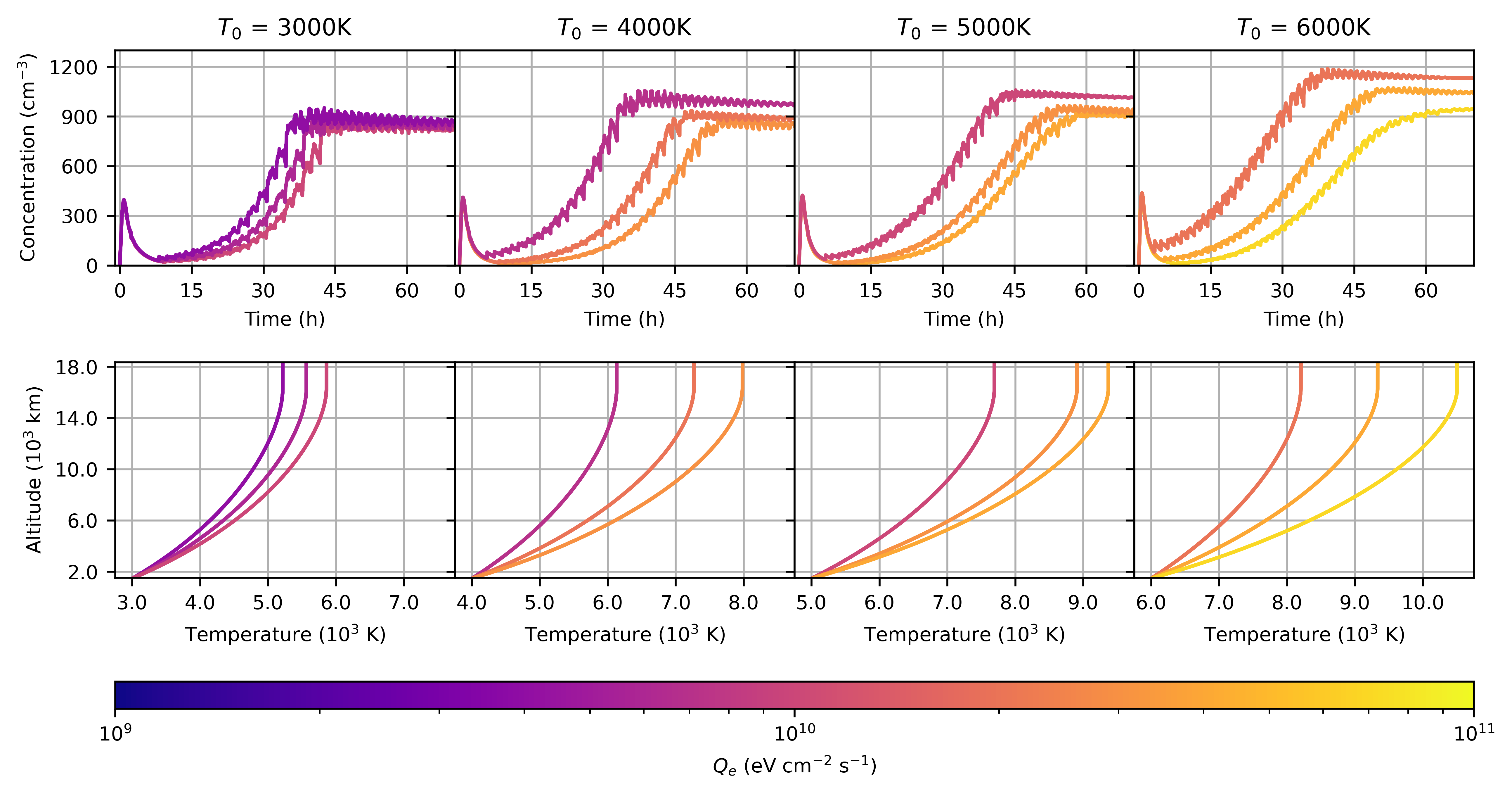}
\caption{(Top): Total equatorial concentration (sum of H$^+$, He$^+$, and O$^+$) vs time with $T_0$ indicated by column and $Q_e$ designated by color. (Bottom): Altitude vs final electron temperature, which is also assumed to be the final ion temperature. These temperature profiles set in within approximately one hour of refilling and persist for the remaining simulation time with minimal variation, as seen in Fig. \ref{SampleRefilling}. The categorization between columns and color is maintained such that per column, lines of the same color in the top and bottom rows are from the same refilling simulation. }
\label{AssortedSimulations}
\end{sidewaysfigure}

    The model's newly developed capability to account for more realistic temperature variations \cite{fitzpatrick_enhanced_2026} enabled a series of simulations with selected combinations of $T_0$ and $Q_e$. Fig. \ref{AssortedSimulations} shows the resulting equatorial refilling (identical to the left panel of Fig. \ref{SampleRefilling}, except the concentrations of each ion are summed together) in the top row and the corresponding final temperature profiles in the row below. A temperature gradient $\nabla \: T$ along the field line can be observed and is a positive quantity assuming the convention of \citeA{comfort_thermal_1996} such that the temperature gradient is measured with increasing altitude. To reduce $\nabla \: T$ to a single value, we also assume temperature varies linearly with altitude as does \citeA{comfort_thermal_1996}. Although this is incorrect (as shown by the right panel of Fig. \ref{SampleRefilling}), the assumption is made for an easier comparison between simulations of different levels of temperature variation across altitude, which result from the varied $T_0$ and $Q_e$. Since each simulation exhibits the same temperature structure (as shown by the bottom row of Fig. \ref{AssortedSimulations}), the calculation of $\nabla \: T$ is primarily indicative of the temperature difference between the highest (approximately 18,000 km) and lowest (1,500 km) altitudes.
    
    The length of each refilling stage for a given $\nabla \: T$ profile can be identified by the same means as was described for Fig. \ref{SampleRefilling}, where refilling starts in the early stage and the particles initially cross the equator, but progresses into the late stage once the steady increase toward the saturated concentration begins. By comparing $\nabla \: T$ with the length of each refilling stage for its column of $T_0$, it can be observed that as $\nabla \: T$ increases, the durations of both early and late-time refilling also increase.

\begin{figure}
\includegraphics[center]{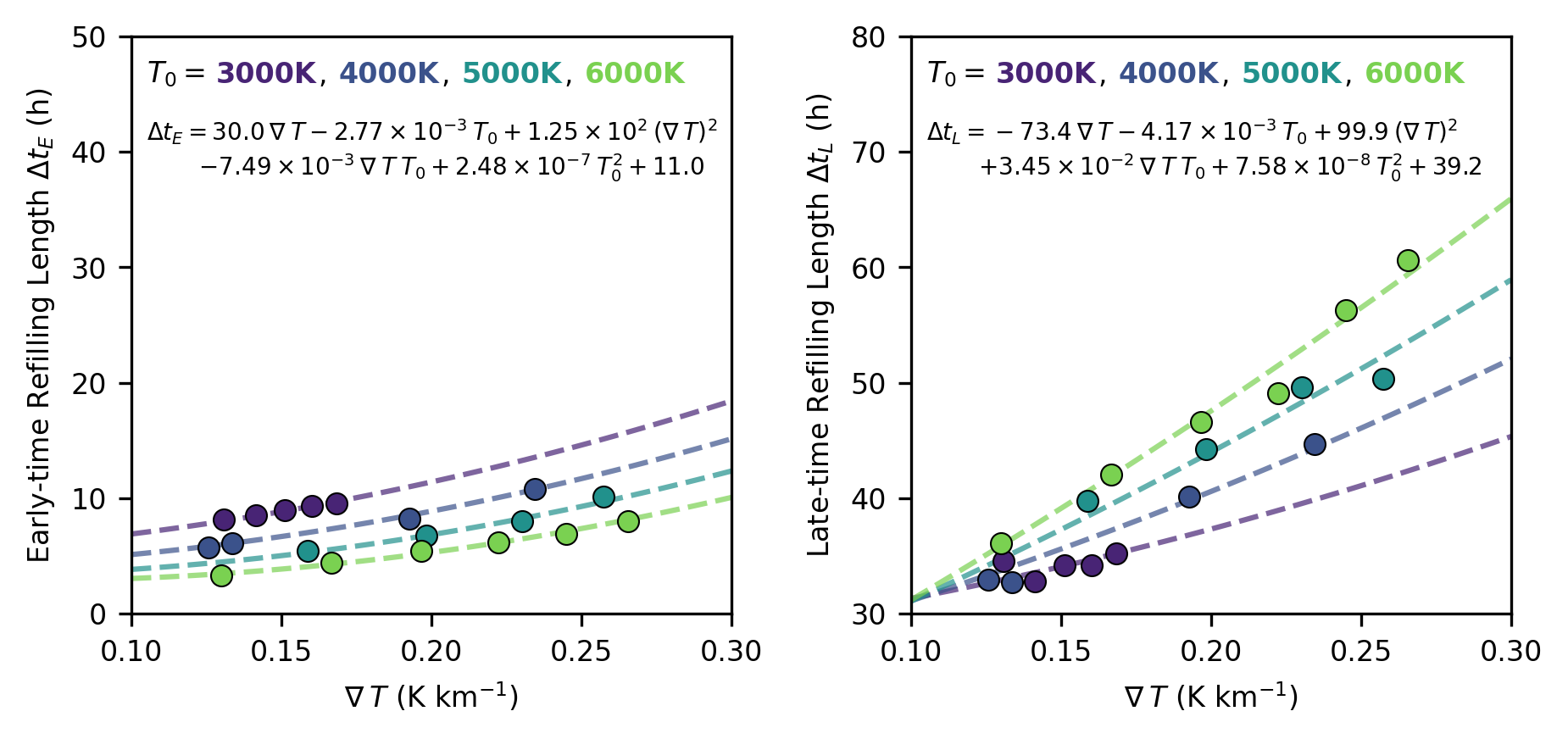}
\caption{(Left) Early-time and (right) late-time refilling length as functions of $\nabla T$ and color designating $T_0$. The circles correspond to results of particular refilling simulations, and the dashed lines are solutions at a given temperature to the multivariate second-degree polynomial regression fit (written out in the top-left of each plot) of the simulation results.}
\label{RegressionFits}
\end{figure}

    A multivariate polynomial regression analysis was conducted to evaluate the dependency of the refilling stage lengths on $\nabla \: T$ and $T_0$ across the 19 considered simulations, as depicted in Fig. 3. The early- and late-time refilling lengths for each simulation are represented by the circles plotted in the left and right panels, respectively.  $\nabla \: T$ is given by the $x$-axis and $T_0$ by color. Performing a second-degree polynomial regression for each early and late-time refilling lengths independently resulted in the fit equations presented in the top-left of each plot. The robustness of the fit is demonstrated by the significantly high correlation coefficients R$^2$ of $99.2\%$ and $98.4\%$ for early- and late-time refilling, respectively. 

    Since the form of the relationship between the refilling length of each stage with $T_0$ and $\nabla \:T$ is unknown, the fit equations in Fig. 3 are the best available means to study how the refilling length depends on $T_0$ and $\nabla \:T$.  The Supporting Information contains contour plots of each term in each fit equation across $T_0$ and $\nabla \:T$ to display the significance of each term's contribution to the overall refilling length with varying combinations of $T_0$ and $\nabla \:T$. For example, the terms for the early-time refilling length fit all contribute similarly, whether by increasing or decreasing, to the total refilling length. On the other hand, for late-time refilling, the degree-two terms are nearly negligible while the cross terms are the most prominent. It also appears that the early- and late-time refilling lengths both extend with increasing $\nabla \: T$, but are oppositely influenced by increasing $T_0$. Further analysis can be done by evaluating the Supporting Information. \par

    The relationship between $\nabla \: T$ and refilling stage lengths is also consistent with the pressure gradient due to a geomagnetic storm eroding the plasmasphere's outer layers. The enhanced pressure gradient further drives ionosphere-outflows of plasma to refill the plasmasphere \cite{darrouzet_earths_2009}. If the pressure of ions and electrons is taken as $P_{i/e} = n_{i/e}kT$ and differentiated across the flux tube, the result is 
\begin{equation} 
\frac{\partial P}{\partial s}=k \left( T\frac{\partial n}{\partial s} + n \frac{\partial T}{\partial s}\right)
\end{equation}
    where $n$ is their respective number density, $k$ the Boltzmann constant, $T$ the space and time-dependent temperature, and $s$ the position coordinate along the flux tube with increasing altitude \cite{chatterjee_development_2018, comfort_thermal_1996}. With this convention of $s$, the conditions at the start of refilling are such that $\frac{\partial n}{\partial s} < 0$ and $\frac{\partial T}{\partial s} > 0$. \par

Although a larger ionospheric outflow is induced by the increased $\frac{\partial P}{\partial s}$ after a geomagnetic storm, the outflow naturally reduces $\frac{\partial P}{\partial s}$ in efforts to return to the quasi-equilibrium state initially disrupted by the geomagnetic storm. The heating and erosion of the storm also cause abnormally large magnitudes of $\frac{\partial n}{\partial s}$ and $\frac{\partial T}{\partial s}$ during refilling. Since $\frac{\partial T}{\partial s}$ becomes a rapidly established constant in the model (as seen in the right panel of Fig. \ref{AssortedSimulations}), reducing $\frac{\partial P}{\partial s}$ requires  $|T \frac{\partial n}{\partial s}|$ to approach $n \frac{\partial T}{\partial s}$. Consequently, a larger $\frac{\partial T}{\partial s}$ requires a larger $|\frac{\partial n}{\partial s}|$ to achieve the same reduction of $\frac{\partial P}{\partial s}$. However, a larger $| \frac{\partial n}{\partial s} |$ inhibits refilling as it enforces a larger difference between the base altitude and equatorial concentrations, resulting in a longer duration of refilling. Thus, it can be concluded that a larger $\frac{\partial T}{\partial s}$ should extend the length of both stages of refilling, as suggested by the results of Fig. \ref{AssortedSimulations}, 3. \par

    Demonstrating that the model's refilling lengths depend on $T_0$ and $\nabla \: T$ implies that these variables could be one of the reasons why two-stage refilling has only been reported by select plasmasphere refilling studies. If temperature and heating conditions are such that early-time refilling is shortened beyond the data's ability to resolve, there would be no apparent distinction between the refilling stages. In Fig. 5 of \citeA{sojka_refilling_1985}, refilling rates derived from geosynchronous orbit data are compared over day-long intervals. The early stage refilling durations in our study are all less than 12 h, and thus would have been averaged-out in \citeA{sojka_refilling_1985}'s type of analysis. Meanwhile, \citeA{lawrence_measurements_1999} and \citeA{su_comprehensive_2001} predetermined early-time refilling lengths to calculate refilling rates across. The length of time a flux tube spends in early-time refilling was designated as the time it spent on the dayside since the start of refilling ($\leq$ 12 h). Since the already-assumed refilling lengths are more comparable to the lengths calculated in our study, the rate of early-time refilling is not overpowered by that of late-time refilling and can thus be observed as its own refilling stage. More recently, \citeA{bishop_global_2024} and \citeA{bishop_superposed_2025} analyzed plasmasphere refilling with the Van Allen Probes. The two probes orbited such that the time between measurements ranged between 4.5-9.0 h. As only 40\% of their refilling events appeared to have two stages of refilling, it is possible that the data of the other 60\% had too limited time resolution to capture early-time refilling, especially if $T_0$ and $\nabla \: T$ in these events further shorten the length of the early stage. 
    
    Nevertheless, a more careful analysis of the spatial and temporal resolution of plasmasphere refilling observations is necessary to confirm whether early-time refilling becoming too short results in ``one-stage" refilling events. Additionally, if other mechanisms are discovered to also condense early-time refilling, they should be compared to evaluate the significance of temperature and heating on shortening the early stage of refilling.

    Recognizing the significance of $\nabla T$ and $T_0$ in two-stage refilling was made possible by this model, but the accompanying simplifications must be noted. The model at its present state remains 1D and it being hydrodynamic limits its accuracy in particularly low concentration conditions. Therefore, the model is more accurate when describing late-time refilling as opposed to early-time refilling. The boundary conditions for concentration and temperature are time independent, thus ignoring time variations in the thermal properties of ionospheric plasma outflow. The value of $Q_e$ in each simulation is similarly constant in time, but also constant in space. Therefore, the equatorial heating of ions via wave particle interactions discussed in \citeA{singh_effects_1990} is not accounted for. This spatially-dependent form of heating results in a mirror force that slows the plasma streams from the ionosphere flowing toward the equator, consequently reducing the duration of refilling.  Another assumption made is the equivalent temperature of electrons and ions, even though electrons are generally at a higher temperature than ions. Hotter electrons can shorten refilling timescales due to an enhanced ambipolar electric field \cite{liemohn_selfconsistent_1997}. Identifying the previously listed assumptions and their impacts on the current analysis provides context for the results when compared to other studies and observations. \par

    Future developments of the model will reduce the number of simplifications and make the refilling results increasingly realistic and thus more easily comparable to observation. Time dependent heating and boundary conditions could be implemented with ionosphere models such as \citeA{maruyama_new_2016}. Even larger-scale time dependence, for instance, across seasons and solar cycle, could be more informative. As the model incorporates more variability, real-life refilling events could be simulated and provide a more direct means of comparison with observation.

\section{Conclusion}

This study analyzes numerical experiments of two-stage plasmasphere refilling to demonstrate how the lengths of early and late stage refilling depend on the field-aligned electron temperature gradient $\nabla \: T$ and the initial/boundary temperature $T_0$. The hydrodynamic model used \cite{chatterjee_multiion_2019, chatterjee_development_2020} exhibits two stages of refilling and was recently enhanced to incorporate spatiotemporally varying electron temperature, allowing comparison between refilling lengths with $\nabla \: T$ and $T_0$. A suite of simulations sampling different values for $T_0$ and $Q_e$ produced refilling events with different $\nabla \: T$. Assuming temperature depends linearly on altitude, a value of $\nabla \: T$ was designated for each simulation. It was observed that for a given value of $T_0$, the lengths of each refilling stage increased as $\nabla \: T$ increased. This conclusion was supported by conducting a second-degree polynomial regression fit, where the refilling lengths of each stage showed a strong dependency on $\nabla \: T$ and $T_0$ with $R^2=99.2\%$ and $98.4\%$ for early and late stages, respectively. A proposed implication of these results is the ability of the temperature structure to shorten early-time refilling to the point that it becomes indistinguishable with late-time refilling, thus becoming a source of apparently one-stage refilling events.

The relationship between refilling length and electron temperature was easier to recognize because of assumptions made by the model, but reducing these will improve the model's ability to depict increasingly realistic refilling scenarios. The currently time-independent boundary conditions and constant heating rate should be replaced by more detailed ionospheric coupling and heating mechanisms for further analysis. Developing the model's physicality allows for more detailed comparison with observation, such as to evaluate whether one stage refilling events are influenced by temperature conditions. As questions continue to arise about plasmasphere refilling and related ionosphere-magnetosphere coupling processes, modeling perspectives like the one taken by this paper are valuable for discerning what underlying physical processes explain observed phenomena.

\section*{Open Research Section}
 The model data used to produce each figure will be provided on CU Scholar (https://scholar.colorado.edu). In the meantime, the data is provided only to the reviewers separately.

\section*{Conflict of Interest declaration}
The authors declare there are no conflicts of interest for this manuscript.

\acknowledgments
We would like to thank the University of Colorado Boulder and the Laboratory for Atmospheric and Space Physics for their support of this work. JF and NM acknowledge Grants NASA 80NSSC20K1817, 80NSSC22K1023, 80NSSC23M0192, 80GSFC23CA004, 80NSSC20K1351, 80NSSC25K7762, NSF Grant AGS-2412296, and AFOSR FA9550-24-1-0013. XC would like to thank Grants NASA 80NSSC23K0096, 80NSSC24K1112, NSF grant AGS-2247255, and AFOSR YIP FA9550-23-1-0359. JB gratefully acknowledges NASA/LWS award  80NSSC22K1023 through subaward 1562616 to UCLA. JG NASA acknowledges LWS grant 80NSSC22K1023, subcontract 1562616 with U Colorado LASP. This work utilized the Alpine high-performance computing resource at the University of Colorado Boulder. Alpine is jointly funded by the University of Colorado Boulder, the University of Colorado Anschutz, Colorado State University, and the National Science Foundation (award 2201538).

%
%

\bibliography{references.bib}

%
%
%
%
%

\end{document}